\def\year{2019}\relax
\documentclass[letterpaper]{article} 
\usepackage{aaai19} 
\usepackage{times} 
\usepackage{helvet} 
\usepackage{courier} 
\usepackage[hyphens]{url} 
\usepackage{graphicx} 
\usepackage{graphicx} 
\usepackage{amsmath}
\usepackage{color}

\newcommand{\citet}[1]{\citeauthor{#1}~\shortcite{#1}}

\title{Using Search Queries to Understand Health Information Needs in Africa}

\author{Rediet Abebe \\ Cornell University
\And
Shawndra Hill \\ Microsoft Research
\And
Jennifer Wortman Vaughan \\ Microsoft Research
\AND
Peter M. Small \\ Rockefeller Foundation
\And
H. Andrew Schwartz \\ Stony Brook University
}

\begin{document}

\maketitle
\begin{abstract}
The lack of comprehensive, high-quality health data in developing nations creates a roadblock for combating the impacts of disease.
One key challenge is understanding health information needs of people.
Without understanding people's everyday concerns, health organizations and policymakers are less able to effectively target education and programming efforts.
In this paper, we propose a bottom-up approach that uses search data to uncover and gain insight into health information needs of individuals in Africa.
We analyze Bing searches related to HIV/AIDS, malaria, and tuberculosis from all 54 African nations.
For each disease, we automatically derive a set of common topics, revealing a widespread interest in various types of information, including disease symptoms, drugs, concerns about breastfeeding, as well as stigma, beliefs in natural cures, and other topics that may be hard to uncover through traditional surveys.
We expose the different patterns that emerge in health information needs by demographic groups (age and gender) and country. Using finer-grained data, we also uncover discrepancies in the quality of content returned by search engines to users by topic and highlight differences in user behavior and satisfaction. Combined, our results suggest that search data can help illuminate health information needs in Africa and inform discussions on health policy and targeted education efforts both on- and off-line.
\end{abstract}

\section{Introduction}
New technologies and data-sources are constantly being leveraged to upgrade and supplement the design, monitoring, and evaluation of health policy in the developed world. There is, however, a substantial gap in the availability and quality of health data between developing and developed nations. In many developing nations, even when health-related information is collected, it is often neither comprehensive nor digitized. A 2014 regional report by the African Union highlights this issue, noting: ``Unless gaps are identified early and accurately, simply providing a raft of general interventions will not meet the real health needs of the people in the Region" \cite{AU}.

This lack of data can be a roadblock to identifying major public health concerns and implementing effective interventions. While targeted education addressing individuals' health needs is a critical tool for combating disease, health organizations and policy makers struggle to identify what knowledge individuals in developing nations seek and whether their health information needs are being met. It is especially urgent to understand how such needs vary by region and demographic groups since the impact of diseases---their prevalence, progression, and transmission rates---as well as people's disease knowledge and attitudes vary regionally and demographically. Limited understanding on the health information needs of individuals hinders the efficacy of gender- and age-specific programming \cite{whogender,ungender,GF,UNDPTb,UNDPMal}.

In this paper, we take a step towards narrowing this gap, focusing on the problem of identifying and measuring people's everyday health information needs, concerns, and misconceptions. We use Bing search queries originating in all 54 African nations to explore which themes related to infectious disease people are most interested in getting information about, as evidenced by their searches. We focus on HIV/AIDS, malaria, and tuberculosis because, together, these three diseases account for 22\% of the disease burden in sub-Saharan Africa~\cite{GBD}.

Search data provide a wealth of information on people's real-time activities, experiences, concerns, and misconceptions relatively cheaply~\cite{book,kern2016methods}, allowing us to obtain potentially hard-to-survey information in a bottom-up manner.
In contrast, most data-driven efforts aimed at mitigating the impact of disease in data-sparse regions, including the Global Burden of Disease Study and the African Health Observatory, have used a top-down approach, actively collecting data with a particular goal in mind~\cite{GBD,AHO}.
Such approaches, while helpful, are often limited in their ability to provide a thorough and comprehensive overview of people's information needs, attitudes, and misconceptions. Existing bottom-up solutions to this problem, such as the West Africa Health Organization's study of health information needs in West Africa, primarily make use of manual interviews~\cite{WAHO}. These approaches can obtain a comprehensive picture of individuals' needs, but are difficult to scale, expensive, and time-consuming. Analyzing search data is a natural candidate for scaling up studies not only because it addresses some of these challenges, but also because search logs have already been shown to contain large quantities of information related to serious conditions in other contexts~\cite{choudhury,book}.
Despite the fact that Internet penetration in Africa is growing rapidly---31\% of the population is currently covered, with nearly 8,500\% growth since 2000~\cite{ITU}---to our knowledge, no prior work has looked specifically at search data to understand health information needs in all African nations.

\vspace{2mm}

\noindent \textbf{The Present Work.} We analyze Bing search data related to HIV/AIDS, malaria, and tuberculosis from all 54 African nations. We uncover themes in which individuals are interested using latent Dirichlet allocation (LDA), a standard generative model for automatically extracting topics from text \cite{blei}. The topics that emerge cover basics such as symptoms, testing, and treatment, as well as hard-to-survey topics such as stigma
and discrimination, beliefs in natural cures and remedies, and concerns about the impact of gender inequality in HIV transmission.
We explore the ways in which the popularity of these topics vary by age,
gender, and location. We expose patterns including that searches related to pregnancy and breastfeeding and relatively more popular among women while searches related to cure news are relatively more popular among men.

Delving into the content returned to users, we compare the organic search results returned for different topics and quantify the discrepancies in the quality of information returned to users. These results highlight unmet health information needs, concentrated misinformation related to specific health topics, and differences in user satisfaction by topic. We discuss the limitations of our approach, including the difficulty of extrapolating our observations to the wider population of Africa, and the danger of overlooking the health concerns of communities who are not on the web. Finally, we highlight potential implications of these analyses on health policy and education efforts both on- and off-line.

\section{Related Work}

\vspace{2mm}

\noindent \textbf{Health Information Needs.} Health information seeking behavior plays a key role in combating the burden of diseases. Online behavior can provide an important lens, especially for stigmatizing conditions (such as STIs), where off-line behavior may be harder to collect~\cite{fox}. Health information is central to disease control; for instance, HIV management requires extensive informational support to maintain the well-being of those affected and their caretakers. However, there is inadequate understanding of individuals' health information seeking behavior, disease knowledge, and perceptions for the three diseases of interest in this paper \cite{hivknowledge,hivcanada}.

This lack of knowledge is especially prominent for individuals living in developing nations. There are relatively few studies, and existing studies are often limited to specific subpopulations. For instance, \citet{ugandahiv} explore the HIV/AIDS knowledge, attitudes, and practices of Ugandan individuals with disabilities. Similarly, \citet{malawihiv} set out to understand how couples living with HIV in Malawi obtain sources of information and reproductive decisions. Studies covering all 54 nations in the continent have often focused on aggregate health outcomes, such as quantifying the burden of diseases by country, rather than health information consumption.


\vspace{2mm}

\noindent \textbf{Search Data for Health.} Search and other large-scale Web data have emerged as key for understanding health patterns and health information consumption \cite{reddit,fox,sillence,liu,eysenbach2002consumers,spink2004study,choudhury}. Online health information seeking behavior is known to be connected to off-line behavior and can inform health policy \cite{fiksdal2014evaluating,ling2016disease,zheluk2013internet,ocampo2013using,hivonline}. Search data is especially valuable as it is real-time, detailed, relevant, and gives less-filtered insights into individuals' health information needs \cite{wwwjama}.

Despite these findings, studies focusing on the use of search engines as a medium for obtaining health information related to the three diseases have remained small scale, often limited to surveys and focused on developed nations \cite{hivonline,hivcanada,shuyler2003patients}. There is need to understand individuals' search behavior before attempting to target relevant information to individuals whether on- or off-line \cite{fiksdal2014evaluating}.

A closely related line of work to ours is surveillance and case finding, where there is extensive work related to HIV/AIDS, malaria, and tuberculosis \cite{zhou2010notifiable,ocampo2013using}. This work shows promising results connecting on- and off-line behavior and suggests that search data can be valuable as an information source for health. Similar work related
to other diseases include using search data for influenza outbreaks
\cite{flu1,flu2,flu3,flu4,flu5}, dengue fever \cite{dengue1,dengue2}, norovirus outbreaks \cite{noro}, bacterial infections \cite{bacteria}, and many other outbreaks \cite{gen1,gen2,gen3,gen4}.

\vspace{2mm}

\noindent \textbf{Impact of Demography on Health.} Health outcomes can vary drastically by demographics, especially in developing nations. Gender and age impact likelihood of being infected and ability to obtain care and treatment. For instance, it is known that 61\% of all sub-Saharan individuals living with HIV are women, and women in the 15--24 age group are three times more likely than men in the same age group to acquire HIV \cite{whohiv}. Men are more likely to develop and die from tuberculosis \cite{UNDPTb}. Pregnant women are disproportionally impacted by malaria \cite{UNDPMal}.

\begin{figure*}[!ht]
\centering
\includegraphics[width = 4.5cm]{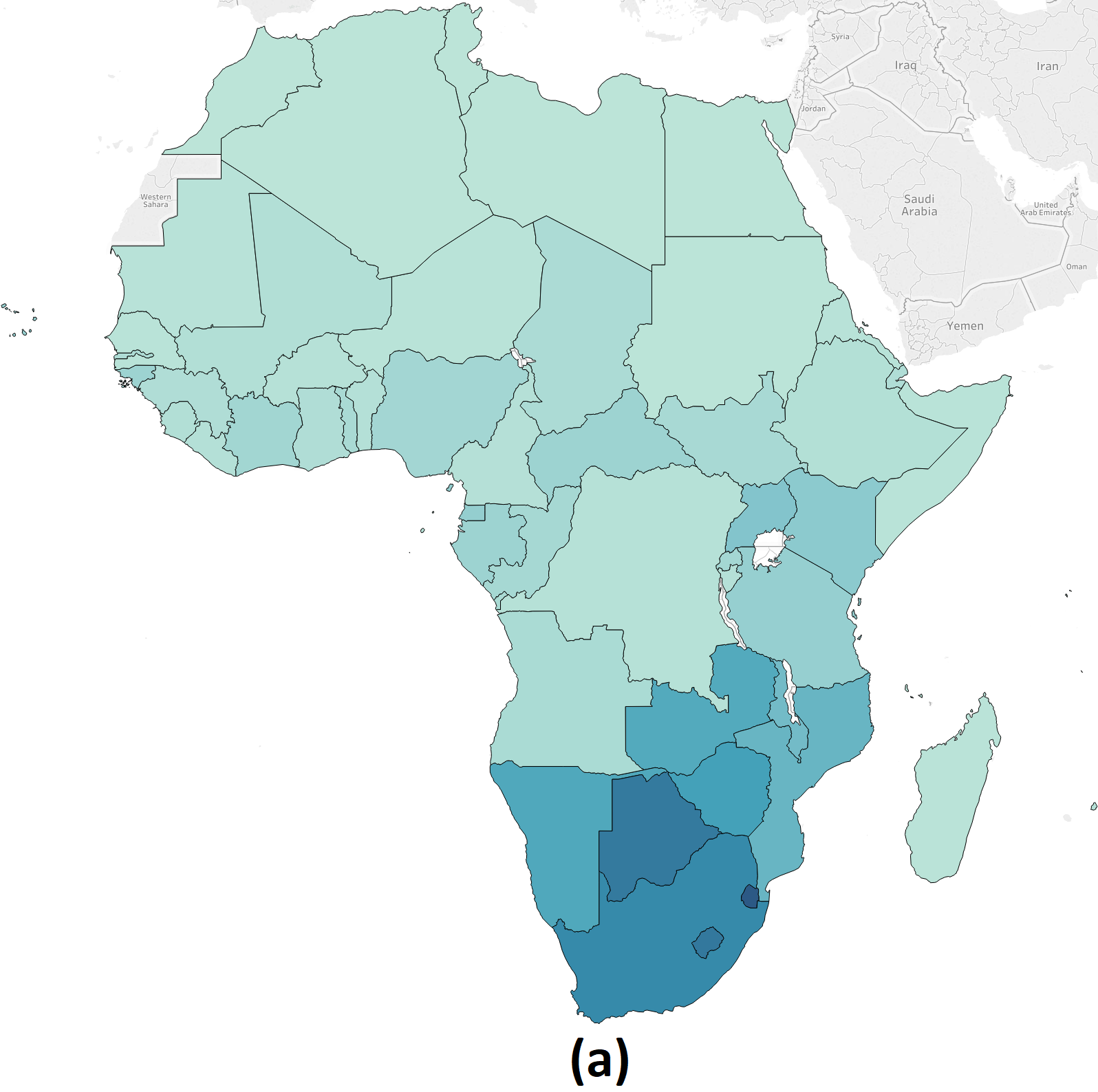}
\hspace{6mm}
\includegraphics[width = 4.5cm]{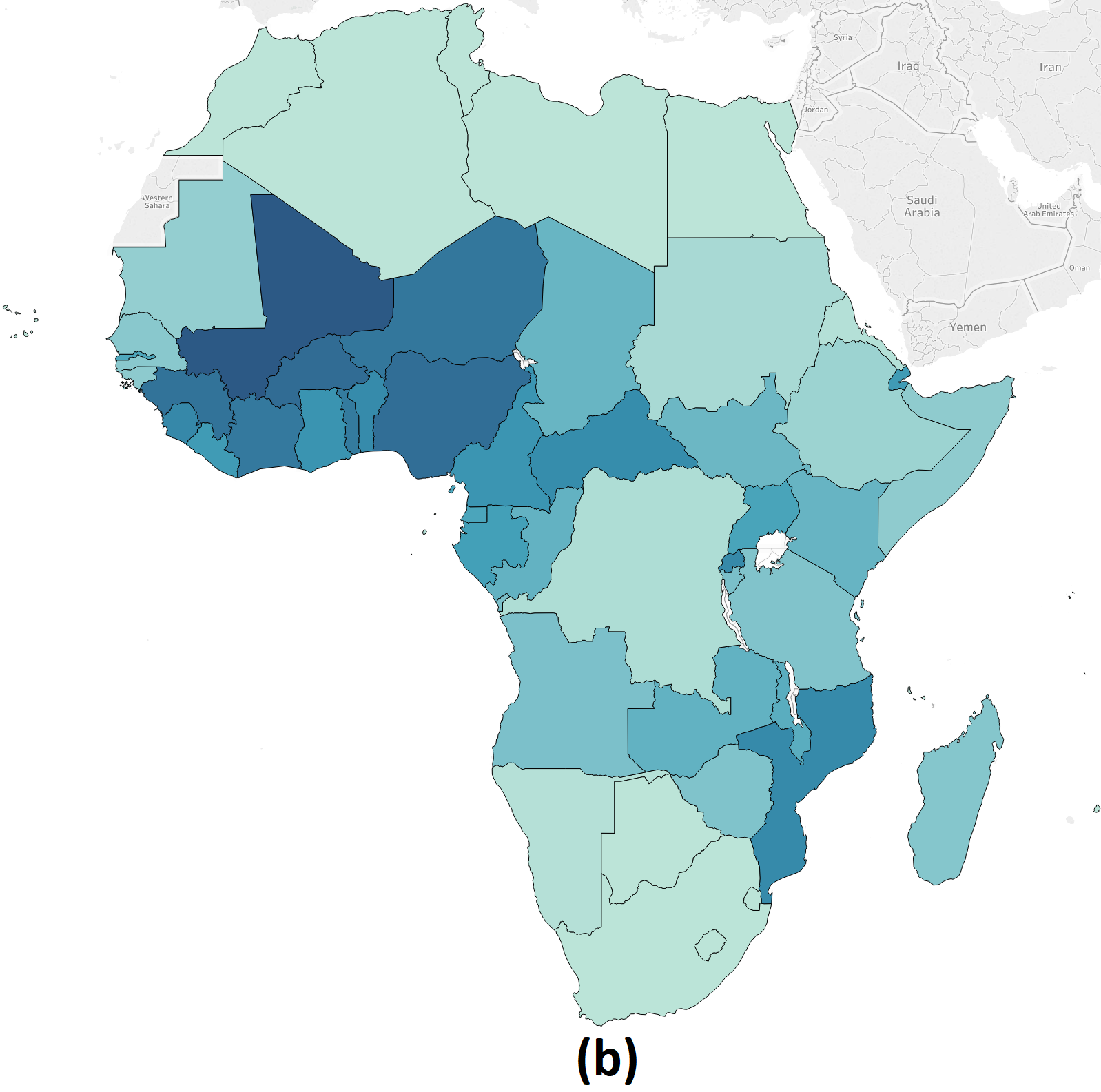}
\hspace{6mm}
\includegraphics[width = 4.5cm]{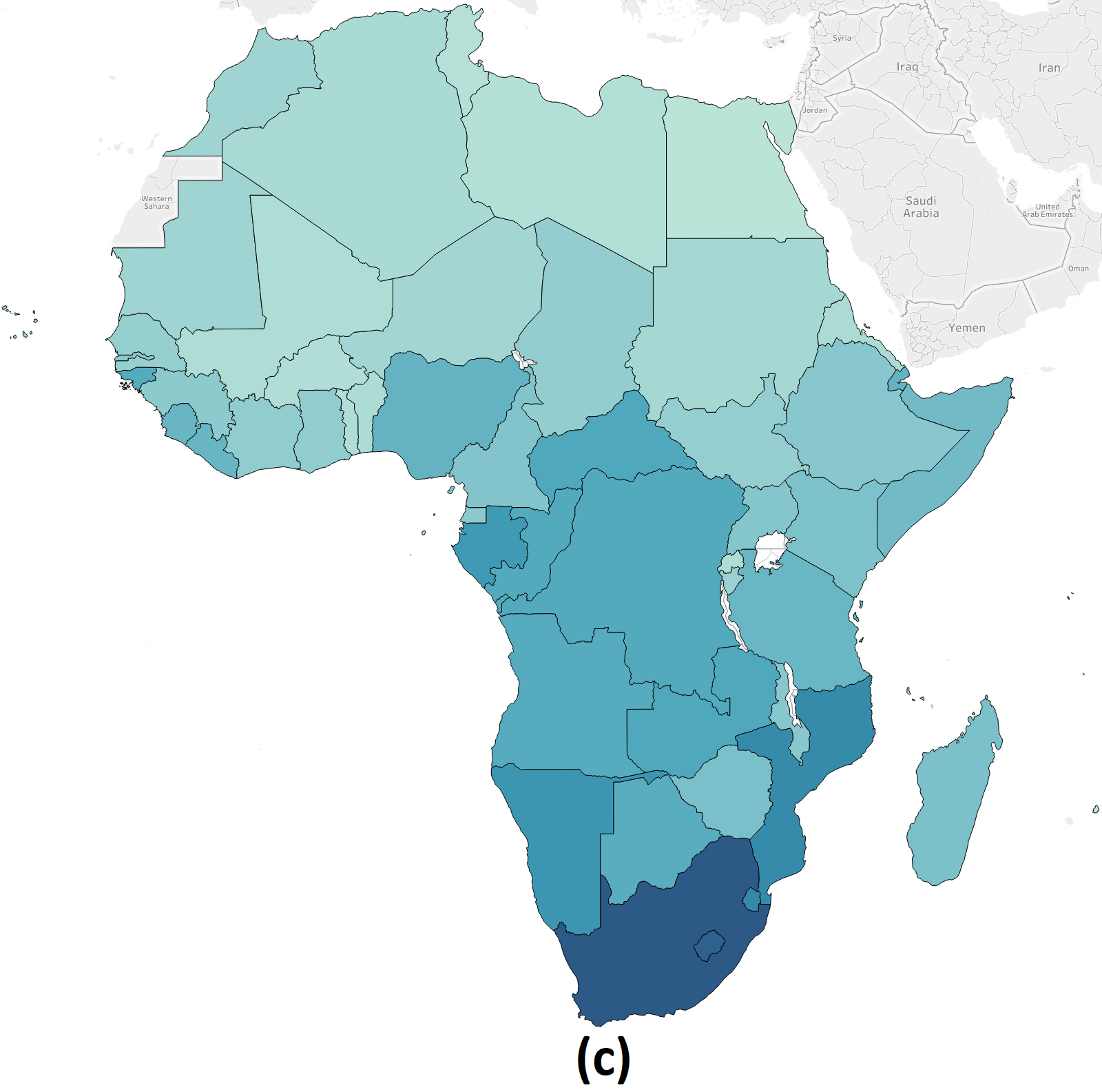}\\
\vspace{2mm}
\includegraphics[width = 4.5cm]{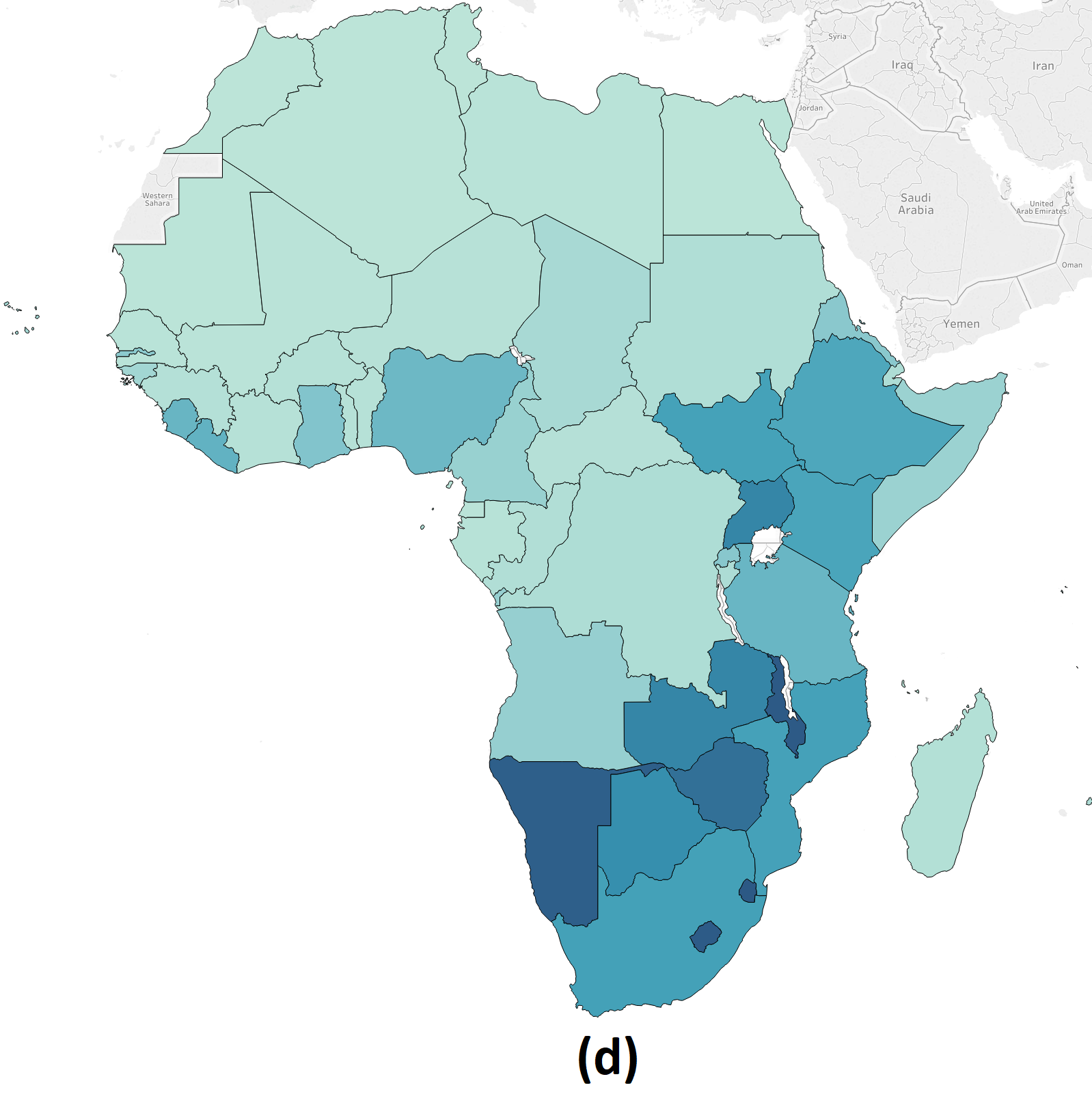}
\hspace{6mm}
\includegraphics[width = 4.5cm]{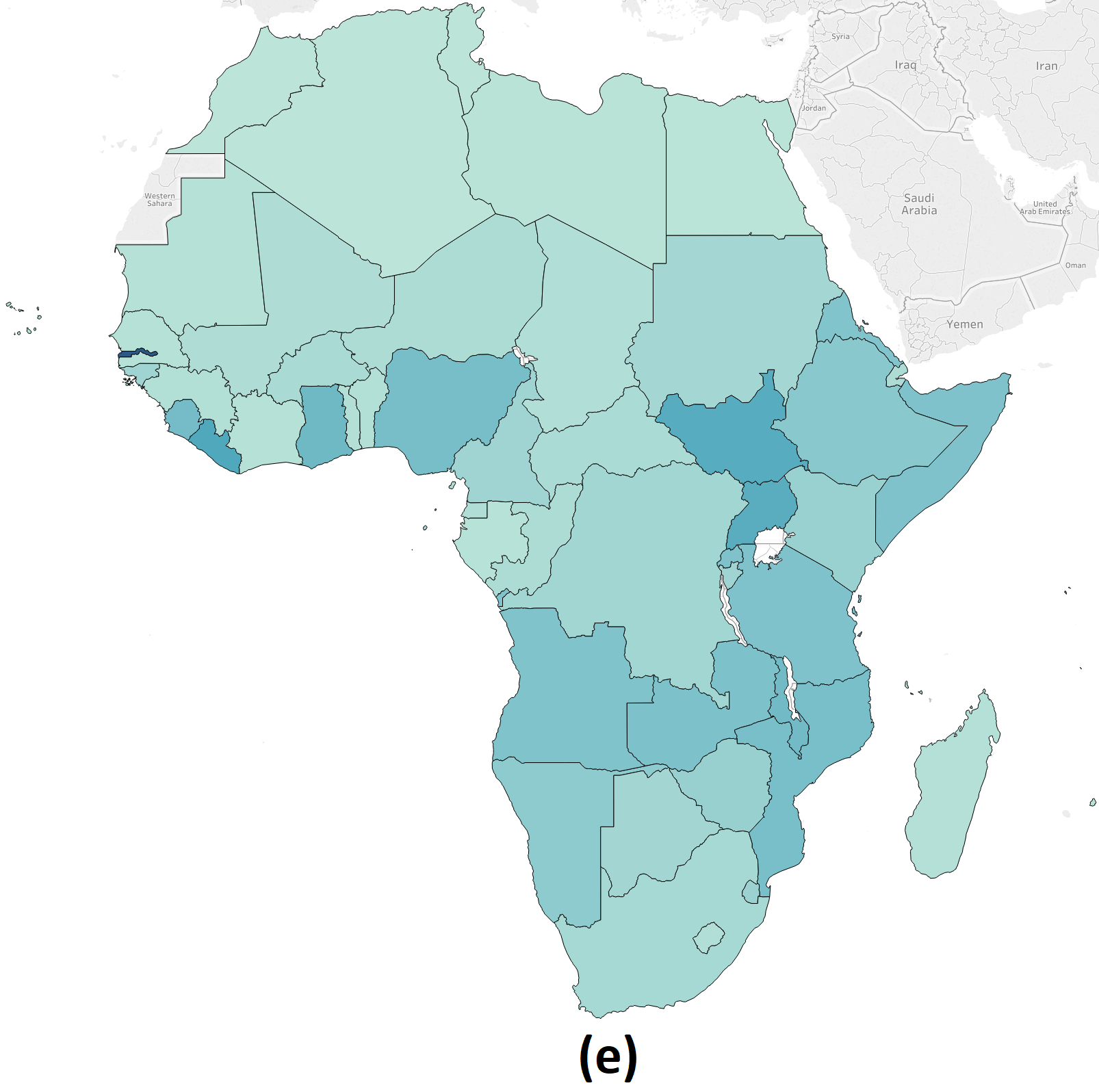}
\hspace{6mm}
\includegraphics[width = 4.5cm]{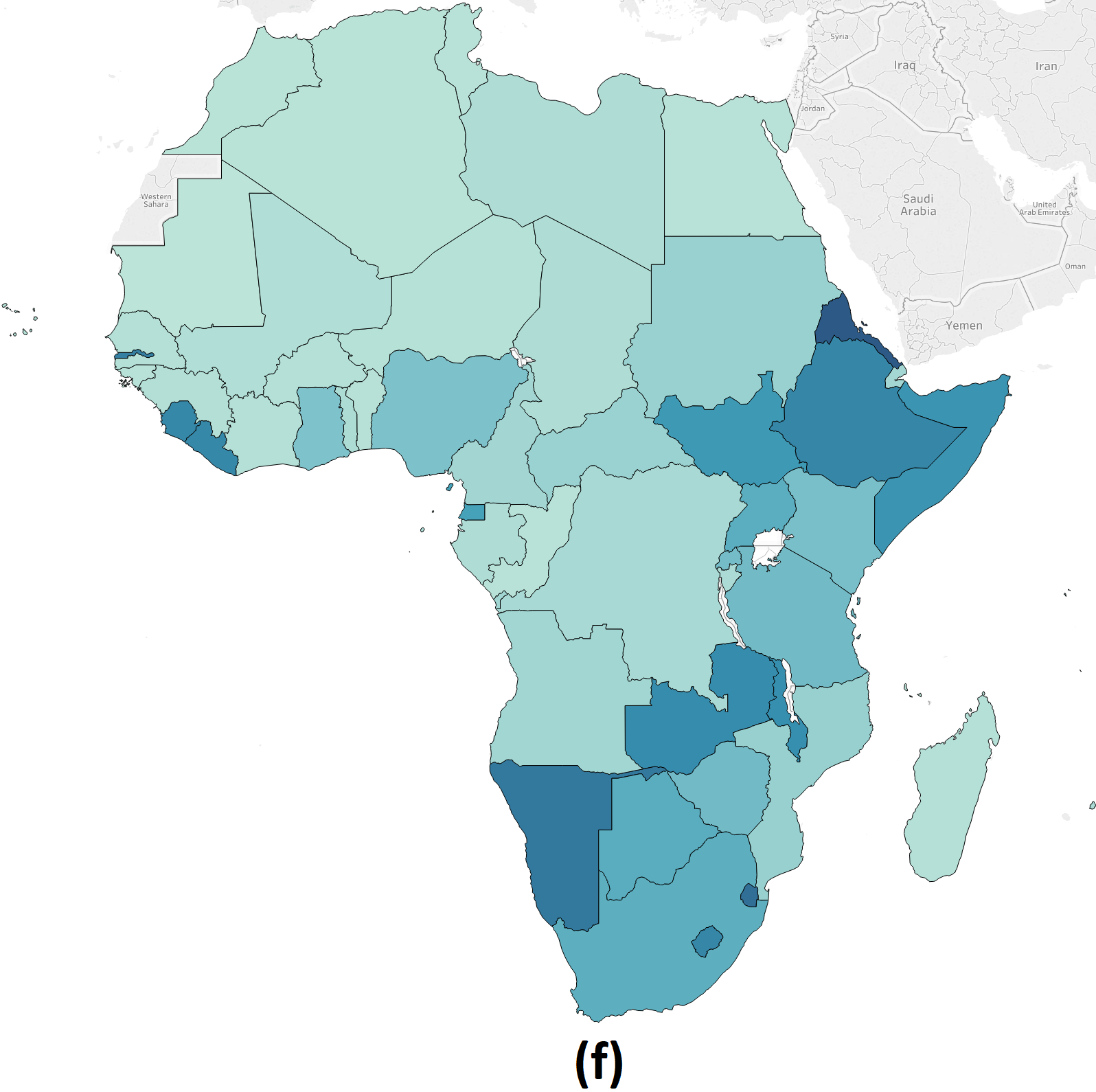}
\caption{
Top: Heat maps showing 2016 rates of (a) HIV/AIDS prevalence (ages 15--49), (b) malaria incidence, and (c) tuberculosis incidence. Bottom: Heat maps showing percentage of total search traffic containing the words (d) ``HIV" or ``AIDS", (e) ``malaria", and (f) ``tb" or ``tuberculosis." \label{fig:searchtraffic}}
\end{figure*}

This variance extends to individuals' knowledge about the diseases of interest \cite{fransen2009young}. Young individuals are evaluated to have incomprehensive knowledge about these diseases, and young women especially so \cite{drhiv,whohiv,chinahiv,kumar,china2}. At the same time, it is difficult to find large age and gender-disaggregated data as they are not routinely collected or reported \cite{UNDPMal,UNDPTb}. This highlights a research gap and potential for computationally-informed policy contributions for effective prevention, coverage, and treatment of these diseases. In resource-constrained settings where funds must be allocated strategically, it is especially prudent to take the varied needs of these groups into account.


\section{Data and Methodology}\label{sec:methods}

To generate the data set of HIV/AIDS queries, we first obtained all Bing search queries containing at least one of the terms ``HIV'' or ``AIDS'' that originated in any of the 54 African nations between January 2016 and June 2017. We consider this time-period since data at the level of granularity described here was available for this time period. Both mobile and desktop searches were retrieved. Each query record in the data consisted of the raw search query, country of origin, and date, along with self-reported age and gender of the user when available. We scrubbed the data to remove HIPAA identifiers: names, addresses, IP addresses, phone numbers, Bing user ids, among others. The data were anonymized for Bing for business purposes prior to the researchers' access. The data sets of malaria and tuberculosis queries were generated in an analogous manner, except with keywords ``malaria'' as well as ``tb'' or ``tuberculosis.''

Figure \ref{fig:searchtraffic} shows two heat maps for each disease. The top maps illustrate the 2016 disease prevalence (for HIV) or incidence (for malaria/tuberculosis) rates for each country, obtained from the World Bank Databank \cite{WB}. The bottom maps illustrate the fraction of total searches made in each country that contain the specific disease terms. There is a high correlation between the fraction of searches about a given disease in a particular country and the disease rate. The Spearman correlation is $\rho=0.714\ [0.689, 0.737]$ for HIV/AIDS, $\rho = 0.402\ [0.360, 0.442]$ for malaria, and $\rho = 0.462\ [0.422, 0.499]$ for tuberculosis. Each correlation coefficient has $p < 0.01$. We view this as reassurance that search queries filtered in this way are pertinent to the diseases in question.

\vspace{2mm}

\noindent \textbf{Disease Topics.} We extracted topics from each data set using LDA, a standard generative statistical model in which each \emph{document} (in our case, an individual search query) is a distribution over \emph{topics}, and each topic is a distribution over words~\cite{blei}. We used the implementation of LDA provided by the Mallet package~\cite{mallet} and the Differential Language Analysis ToolKit as an interface to Mallet for further analysis \cite{dlatk,dlatk2}. We retained all default parameters, with the exception of $\alpha$, the prior on the per-document topic distribution, which we set to $2$ since search queries are shorter than the documents for which LDA is typically used.

\begin{table*}[!htbp]
\caption{Sample LDA Topics for HIV/AIDS, Malaria, and Tuberculosis with Representative Words and Sample Queries}
\begin{tabular}{p{0.05\textwidth}|p{0.1\textwidth}|p{0.5\textwidth}|p{0.25\textwidth}}
\hline
Disease & Topic & 20 Most Representative Words & Sample Queries from Top 100\\
\hline
\hline
& \emph{Symptoms} \newline (2.28\%) & pain, sign, lymph, swollen, nodes, sore, symptom, symptoms, throat, infection, body, back, positive, pains, stomach, fever, neck, headache, glands, patient
& hiv painfull jaw \newline hiv swollen lymph nodes \newline hiv swollen gland throat
\\
\cline{2-4}
& \emph{Natural Cure} \newline (0.74\%) & cure, oil, black, healing, heal, healed, seed, herbs, natural, cures, moringa, kill, cured, testimonials, coconut, traditional, god, garlic, lemon, aloe
& prophet bushiri hiv miracles \newline hiv garlic lemon honey \newline coloidal silver hiv testimonials 
\\
\cline{2-4}
& \emph{Epidemiology} \newline (0.59\%) & statistics, report, 2015, global, unaids, 2016, united, epidemic, besigye, kizza, children, 2014, progress, 2010, response, nations, nigeria, prevalence, million, sa
& unaids global aids report \newline mia khalifa hiv \newline hiv 2030
\\
\cline{2-4}
HIV/ \newline AIDS & \emph{Drugs} \newline (0.85\%) & drug, treatment, patients, abuse, therapy, drugs, resistance, antiretroviral, substance, adherence, alcohol, art, failure, spread, leads, patient, relationship, transmission, effect
& stanford hiv drug resistance \newline hiv drug therapy resistance \newline virological failure hiv 
\\
\cline{2-4}
& \emph{Breastfeeding} \newline (0.66\%) & positive, baby, mother, breastfeeding, breast, mothers, child, born, feeding, babies, birth, give, breastfeed, infant, infected, feed, milk, pregnant, safe, exposed
& hiv exclusive fomular feeding \newline exclusive breast feeding and hiv \newline hiv mom can breast feed baby 
\\
\cline{2-4}
& \emph{Stigma} \newline (0.46\%) & stigma, issues, discrimination, related, ethical, legal, prevention, safety, pdf, workplace, relating, precaution, work, dies, surrounding, universal, reduce, address
& 
hiv aids ethical dilema \newline safty issues relating to hiv-aids \newline aids stigma in garissa
\\
\hline
\hline
& \emph{Symptoms} \newline (0.93\%) & pregnancy, effects, symptoms, early, pathophysiology, treatment, management, effect, pdf, complications, mouth, disease, sign, sore, bitter, throat, nigeria, symptom, dar
& malaria lip sores \newline malaria blisters on lips \newline bitterness in mouth and malaria 
\\
\cline{2-4}
& \emph{Natural Cure} \newline (1.02\%) & cure, natural, home, treat, treatment, remedy, remedies, fever, typhoid, treating, herbal, herbs, medicine, leaf, leaves, good, cures, naturally, lemon
& pawpaw leave malaria remedy \newline papaya leaf malaria 
\newline lipton tea for malaria
\\
\cline{2-4}
& \emph{Epidemiology} \newline (17.39\%) & disease, people, year, africa, deaths, download, die, communicable, nigeria, song, number, virus, cases, died, million, caused, mp3, tropical, soty
& malaria free sri lanka \newline lyrics malaria theme song \newline stoy- malaria mp3
\\
\cline{2-4}
Malaria & \emph{Drugs} \newline (1.31\%) & prophylaxis, treatment, quinine, dosage, pregnancy, dose, cdc, doxycycline, prevention, artesunate, children, malarone, chloroquine, severe, table, treat, guidelines, fansidar, treating
& fansidar malaria dose 
\newline quinine maximum dose malaria \newline artefan malaria dose
\\
\cline{2-4}
& \emph{Breastfeeding} \newline (1.05\%) & drug, drugs, anti, baby, treat, mother, treatment, breastfeeding, medicine, pregnancy, cancer, fight, good, child, taking, months, affect, medication, babies
& malaria breast milk 
\newline can a breast feeding mother take malaria drugs
\\
\cline{2-4}
& \emph{Diagnosis} \newline (1.24\%) & parasite, blood, test, parasites, film, smear, thick, stain, slide, thin, microscope, giemsa, procedure, images, staining, field, medicine, count, density
& dar es salaam malaria \newline malaria swamp \newline swollen lip and malaria 
\\
\hline
\hline
& \emph{Symptoms} \newline (1.80\%)& symptoms, signs, early, stages, warning, list, sign, infection, pulmonary, symtoms, babies, children, infants, symptons, kids, toddlers, cough, baby, symtomps
& night sweat in tuberculosis \newline tuberculosis dry cough \newline tb feet and face swelling 
\\
\cline{2-4}
& \emph{Natural Cure} \newline (0.85\%) & cure, treat, home, treatment, natural, history, remedies, medicine, mdr, group, patient, taboola, utm$\_$source, disease, long, rememdy, traditional, utm$\_$campaign, treatments, herbs
& tuberculosis cure discovered \newline does moringa seed cure tb \newline tb reducing natural remedies 
\\
\cline{2-4}
& \emph{Epidemiology} \newline (0.93\%) & africa, south, statistics, deaths, 2010, death, provinces, sa, stats, show, rate, prevalence, province, 2016, incidence, african, graph, 2015, showing
& tb death toll sa \newline tuberculosis graphs \newline tb death provincial statistic
\\
\cline{2-4}
TB & \emph{Drug} \newline \emph{Side-Effects} \newline (1.44\%) & drugs, effects, side, treatment, anti, medication, effect, drug, line, medications, liver, list, anti-tb, pregnancy, dosage, anti-tuberculosis, patients, adverse, induced
& anti-tuberculosis drugs \newline anti tuberculosis combination \newline 2nd line anti tb\\
\cline{2-4}
& \emph{Diagnosis} \newline (1.28\%) & diagnosis, culture, sputum, mycobacterium, gene, test, laboratory, genexpert, smear, xpert, testing, lab, microscopy, expert, negative, stain, diagnostic, procedure, pulmonary, collection
& tb auramine staining \newline tb culture sensitivity \newline mycobacterium tuberculosis acid-fast stain
\\
\cline{2-4}
& \emph{Drug} \newline \emph{Resistance} \newline (1.11\%) & drug, resistant, resistance, multidrug, treatment, multi, management, therapy, drugs, multiple, pdf, multi-drug, mycobacterium, patients, antibiotic, mdr, active, rifampicin, latent
& multdrug resistant tuberculosis
\newline extensively drug resistant vs multidrug resistant tuberculosis \\
\hline
\end{tabular}\label{table:topics}
\end{table*}

The number of topics extracted is a free parameter that can be tuned. Choosing a large number of topics leads to the discovery of highly specific topics that overlap in theme, while choosing a small number leads to general, multi-theme, and difficult to interpret topics \cite{schwartz2015datadriven}. Before running our analyses, we ran LDA on each data set with different numbers of topics (10, 20, 50, 100, 200, 500, 1,000, and 2,000). Based on a manual inspection of the interpretability and coherence of topics by a health expert, we chose 100 topics for the HIV/AIDS data set and 50 each for the malaria and tuberculosis data sets, which are smaller than the HIV/AIDS data set. These topics were then labeled by a health expert and manually inspected by the authors and checked for any overlooked HIPAA identifiers.

We would ideally like to define representative queries as queries with high weight for the topic. However, the existence of rare words or strings (such as obscure URLs) in a query can result in a query having an artificially high weight for a given topic (abnormally high probability of belonging to a single topic). We thus excluded words that appear fewer than 10 times in the data set. For the same reason, we removed all queries with two or fewer words since these often contained similar issues. (Note that at least one of these must be the name of the disease.)

Additional methods are described below alongside the corresponding results. All statistical significance tests were conducted correcting for false discovery rate with $\alpha = .05$ of testing either 50 or 100 topics using the Benjamini-Hochberg procedure~\cite{benjamini1995controlling}.

\section{Analysis and Results}\label{sec:results}

Our analyses reveal a rich set of themes. The topics output by LDA range from those about standard health information, such as \textit{Symptoms}, \textit{Drugs}, and \textit{Epidemiology}, to those about hard-to-survey concerns, such as \textit{Stigma} and \textit{Natural Cure}. Table \ref{table:topics} shows six sample topics for each disease extracted from the data by LDA, which were hand-chosen and labeled by a health expert to illustrate both the breadth of themes that emerged from the analysis as well as the coverage of hard-to-survey topics. For the HIV/AIDS data set, the full list of topics additionally includes themes such as \emph{Transmission}, \emph{Testing Kits}, \emph{Testing Clinics}, \emph{Gender Inequality}, \emph{Healthy Lifestyle}, \emph{Disease Progression}, and \emph{Celebrity Gossip}. Likewise, the malaria and tuberculosis data sets display a rich set of themes ranging from \emph{Patient Care}, and \emph{Pregnancy}, to \emph{National Programs}.

The second column in Table \ref{table:topics} provides a label for each topic along with a measure of the frequency with which the topic occurs in the data. Specifically, given $\theta_{\text{query}} = p( \text{topic} | \text{query})$, {{the values in parentheses correspond to $\sum_{q \in \text{query}} \theta_{q} \times p(q)$ over all queries, expressed as a percentage}}. This corresponds to the popularity of the given topic.
Some themes, such as breastfeeding, are captured by more than one topic, so the overall frequency with which these themes occur in the data is higher than the numbers suggest. The third column presents the 20 most representative words, according to posterior probabilities of word given topic.
The final column shows randomly selected queries from among the 100 most closely related to the topic.
We show a random sample of queries since the top few most highly ranked queries often differ by only one letter or word.
All typos in the queries are unaltered.

\subsection{Topic Prevalence by Region and Demographics}

We explore whether health information needs, manifested as search queries, vary by country and user demographics. Due to space limitations, we only present results for queries related to HIV/AIDS. The corresponding results for malaria and tuberculosis can be found in the appendix.

For each of the six HIV/AIDS topics listed in Table \ref{table:topics}, we estimate the number of times an HIV/AIDS topic is queried relative to the overall number of queries for HIV/AIDS for the country. We call this quantity topic prevalence and denote it by
$\text{prevalence}(\text{topic} | \text{country})$, which is a measure of the frequency with which a topic is mentioned in queries from a given country. To estimate the prevalence, we need two values---the frequency the topic is searched for in the country and the overall number of searches for any topics in the country. We do not have the exact number of times a topic is mentioned, but we can estimate the frequency with which the topic is used by utilizing the words associated with a topic using
the posterior probabilities, derived from LDA, of a topic given a word, $p(\text{topic} | \text{word})$. We then combine these estimated counts with the relative frequencies of a word given a country, $\text{frequency}(\text{word} | \text{country})$ to get the overall prevalence of the topic in the country:
$$
\sum_{\text{word} \in \text{country}} p(\text{topic} | \text{word} ) \times \text{frequency}(\text{word}| \text{country})
$$

Here, $\text{frequency}(\text{word} | \text{country})$ is derived from maximum likelihood estimation given all words used in queries from the country: the ratio of the number of times a word appears and the total count of all words for the the country.\footnote{Since we use an estimate of the number of times a topic is searched for, we provide results for estimating frequency differently, namely by considering only the top 50 words associated with a topic ranked by $p(\text{topic} | \text{word})$, in the appendix. We obtain qualitatively similar results when we consider all words or the top words to estimate the frequency the topic is searched for in a country.}

\begin{figure}[!ht]
\centering
\includegraphics[scale=0.09]{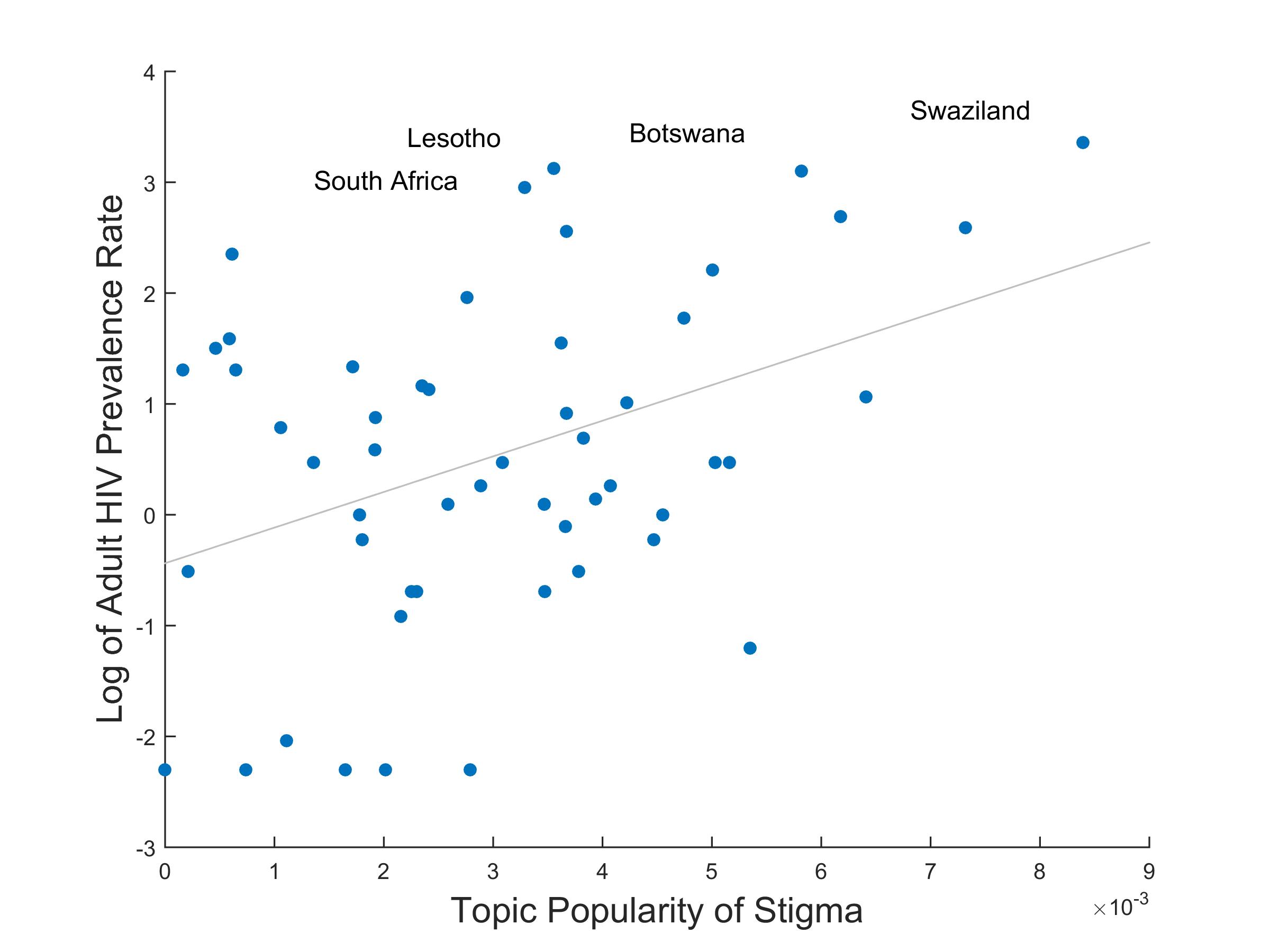}.
\caption{Comparison between topic popularity of \emph{Stigma} in each country with the log of the 2016 adult HIV prevalence rate. Countries with higher topic popularity for \emph{Stigma} tend to have higher HIV prevalence rates.\label{fig:stigma}}
\end{figure}

To explore the association of topic prevalence with HIV prevalence rates across countries, we ran a linear regression using the prevalence values for each of the 100 topics within a given country as the explanatory variables and the 2016 HIV prevalence rate in that country as the dependent variable. Of the six topics listed in Table~\ref{table:topics}, we found a significant 
relationship between the \emph{Stigma} topic and the HIV prevalence rate ($r = 0.473$; multi-test corrected $p < 0.01$), as illustrated in Figure \ref{fig:stigma}.

The observation that the popularity of the \emph{Stigma} topic is correlated with HIV prevalence is consistent with findings from the public health literature.
In particular, smaller-scale studies (often based on survey data), have shown that HIV-related stigma can lead to more risky behavior, lower testing rates, and decreased adherence to antiretroviral therapy, all of which increase transmission rates~\cite{stigma,stigma2}.
An outlier in Figure~\ref{fig:stigma}, \emph{Stigma} is nearly twice as frequent in Botswana as in Lesotho,
despite the two countries having comparable HIV prevalence rates (22.2 versus 22.7). This is consistent with literature indicating that Botswana has struggled with discrimination issues including proposals for mandatory HIV testing \cite{botswanastigma}. Similarly, Swaziland is also known to have one of the highest rates of HIV prevalence, as well as stigma surrounding the disease \cite{swazistigma}.
Further, despite high HIV prevalence rates, the level of stigma has been declining in South Africa, in part due to targeted HIV destigmatization efforts \cite{sastigma}. 

Other topics also exhibit variance in popularity by country. To explore this, we ran a logistic regression for each country, with the standardized topic weights (distribution) of a given search query as the explanatory variables and a binary dependent variable indicating whether or not the query originated in that country. There is notable contrast in the popularity of queries associated with the \emph{Natural Cure} topic across the countries. For instance, the \emph{Natural Cure} topic is popular in Malawi ($\beta = 0.05; p < 0.05$), which has a relatively high HIV prevalence rate of $10.6$, and in Botswana ($\beta = 0.07; p <0.01$), which has an HIV prevalence rate of $22.2$. In Mozambique, which has an HIV prevalence rate of $11.5$, the popularity of the \emph{Natural Cure} topic is relatively low ($\beta = -0.07; p <0.01$).

To explore topic popularity by gender, we ran a logistic regression with the topic weights of a given search query as the explanatory variables and the self-reported gender of the user as the dependent variable, limiting ourselves to those queries for which user demographic information was available. Similarly, to explore topic popularity by age group, we ran an ordinary least squares linear regression, again with the topic weights of a given search query as the explanatory variables, but now with users' self-reported age group as the dependent variable. 
We then ordered the topics by their respective correlation coefficients. The fifteen words with the highest weight from the top ranked topic for each group are shown in Table~\ref{table:demographics}.

Our analysis reveals that topics related to news on HIV/AIDS cures are more popular among men, as well as the 35--49 age group. Topics related to breastfeeding, pregnancy, and family care are more popular among women. For the 18--24 and 25--34 age groups, topics related to symptoms are more popular. Among the former group, topics related to the socioeconomic implications of HIV/AIDS, such as gender inequality, are more popular, while topics related to concerns about transmission to partner and child are more popular among the 25--34 age group.\footnote{Some themes appear in more than one topic. When we identify a novel pattern relating a particular topic to demographics or location in the data (e.g., ``breastfeeding is more popular among women"), we confirmed the same pattern exists for the most similar topics. To measure similarity between topics, we calculate the pairwise hamming distance between topics using the top 20 most representative words. If the observed pattern does not hold across similar topics, we omit the pattern from our positive findings.}

\begin{table}
\caption{Popular topics by age and gender. \label{table:demographics}}
\fbox{\begin{minipage}{23em}
\textbf{Ages 18--24} \emph{(0.083)}: symptoms, signs, early, women, men, infection, stages, months, symptoms, earliest, children, major, syptoms, systoms, rare 
\vspace{1.2mm}

\textbf{Ages 25--34} \emph{(0.070)}: positive, negative, partner, person, man, sex, woman, tested, im, infected, pregnant, baby, husband, wife, infect 
\vspace{1.2mm}

\textbf{Ages 35--49} \emph{(0.063)}: cure, latest, news, research, treatment, discovery, today, vaccine, update, 2015, 2016, 2017, google, breakthrough, recent
\end{minipage}}\\
\vspace{2mm}

\fbox{\begin{minipage}{23em}
\textbf{Women} \emph{(0.115)}: positive, baby, mother, breastfeeding, breast, mothers, child, born, feeding, babies, birth, given, breastfeed, infant, infected 
\vspace{1.2mm}

\textbf{Men} \emph{(0.133)}: cure, news, 2016, latest, vaccine, breaking, 2017, www, development, today, headline, updates, breakthrough, found, feb 
\end{minipage}}
\end{table}

\vspace{2mm}

Finally, we looked at the topic popularity of the six topics of interest from Table \ref{table:topics}. Table \ref{table:topicpopularity} lists the correlation coefficients, where $**$ indicates a p-value of less than 0.01 and $*$ indicates a p-value of less than 0.05.
\begin{table}[!ht]
\caption{Topic Popularity by User Demographics}
\begin{tabular}{p{0.0765\textwidth}|p{0.072\textwidth}|p{0.072\textwidth}|p{0.072\textwidth}|p{0.065\textwidth}}
\hline
& Women & Ages 18--24 & Ages 25--34 & Ages 35--49 \\
\hline
\hline
Sympt. & -0.052** & 0.000 & -0.019* & -0.018* \\ 
Natl Cure & -0.010 & -0.050** & -0.018* & 0.043** \\
Epid. & -0.052** & -0.080** & -0.019* & 0.019* \\
Drugs & -0.016 & -0.020** & -0.041** & 0.030** \\ 
Breastf. & 0.115** & -0.031** & 0.061** & -0.008 \\ 
Stigma & 0.025** & 0.032** & -0.047** & 0.004 \\ 
\hline
\end{tabular}\label{table:topicpopularity}
\end{table}

We again confirm \emph{Breastfeeding} has a higher correlation coefficient for women than for men and for the 25--34 age group compared to the other age groups. Less expected, women and users aged 18--24 are more interested in \emph{Stigma} compared to their demographic counterparts. \emph{Natural Cure} has the highest popularity among the oldest age group (35--49), and the lowest among the youngest age group (18--24). Despite expressing higher interest in \emph{Natural Cure}, the 35--49 age group also has more interest in \emph{Drugs} compared to the other age groups.

\subsection{User Behavior and Quality of Results}

We examine whether user behavior and the quality of search results returned vary across different topics. Differences here would highlight unmet health information needs, concentration of misinformation related to specific topics, and differences in user satisfaction by topic.

We used an expanded version of our HIV/AIDS data set consisting of only those queries that were made during June 2017. In addition to raw queries, country, and search date, this data set contains a list of the first 10 organic web pages returned to the user for each query. It also contains information about which web pages the user clicked on, the amount of time spent on each web page, and the total time spent on the \emph{results page}, the page containing the ten initial links presented to a user after entering a search query.

To compare user behavior across topics, we focused on several standard metrics from the information retrieval literature~\cite{ir}. \emph{Dwell time} measures the total amount of time a user spends looking at the results page and any links that are followed. \emph{Click count} is the total number of links on which a user clicks.\footnote{We also examined \emph{successful click count}, which measures the number of pages a user clicks on with dwell time at least 30 seconds, and \emph{maximum dwell time}, which measures the maximum amount of time spent on a web page. Results for these are similar to the click count and dwell time results, respectively.} Note that these metrics can be used to measure various properties related to user engagement and satisfaction. For instance, dwell time can be used to measure both interest in a web page and ease-of-use, depending on the context and intent of the search query. In our study, we use these metrics to measure user activity. Specifically, we are interested in measuring whether there is variance in user activity by topic as measured by these metrics.

\begin{figure*}[!ht]\label{fig:usersat}
\centering
\includegraphics[width = 5cm]{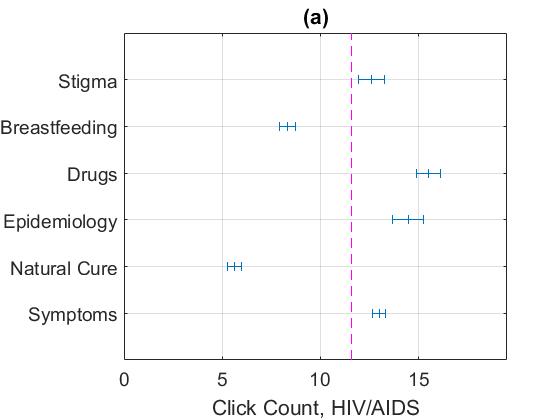}
\includegraphics[width = 5cm]{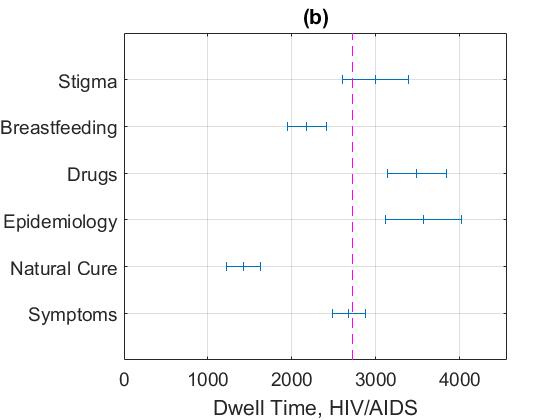}
\includegraphics[width = 5cm]{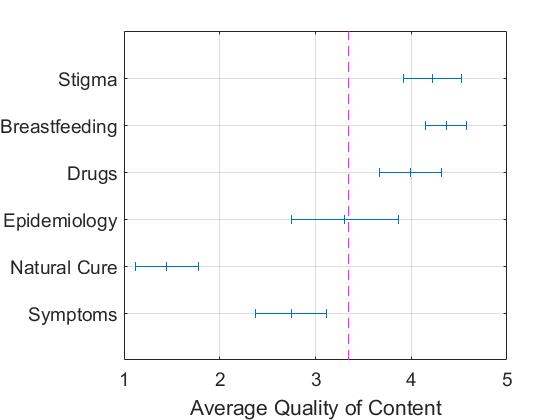}
\caption{Average (a) click count and (b) total dwell time for queries associated with HIV/AIDS topics of interest. The vertical lines represent the mean values across topics. (c) denotes the average quality of content of web pages returned to users for the 30 queries most strongly associated with each HIV/AIDS topic of interest. \label{fig:usersat}}
\end{figure*}

Plots (a) and (b) in Figure \ref{fig:usersat} show how these metrics vary by topic. Both dwell time and click count are significantly lower for the \emph{Natural Cure} topic compared with the other topics of interest. That is, on average, users issuing queries related to the \emph{Natural Cure} topic spend less time exploring the results page and click on fewer links. There are many reasons why this may be the case. It could simply be selection bias---perhaps different types of users search for queries related to the \emph{Natural Cure} topic compared with \emph{Epidemiology}, \emph{Drugs}, or other topics. It could be that users seeking information related to the \emph{Natural Cure} topic find the information they are seeking faster. Another possibility is that the quality of information returned could vary by query.

To examine variance in the quality of content returned, for each of the topics in Table~\ref{table:topics}, we extracted the first link returned to the user at the top of the web results page for each of the 30 queries most strongly associated with the topic. We consider only distinct user/query pairs, which means we ignored duplicate queries from the same user. Each resulting link was independently evaluated for quality (described in terms of relevance, accuracy, and objectiveness, as is standard in information retrieval~\cite{qualityofcontent}) and was ranked by three research assistants on a scale of 1 to 5, with higher values indicating better quality. The research assistants who provided rankings have graduate-level training in medicine or public health, and each website was evaluated by at least one research assistant specializing in the disease of interest. We took the average of these three ratings as the rating for a web page.



Plot (c) in Figure \ref{fig:usersat} shows the average rating across all links and all raters for each topic. On average, the quality of links returned for queries related to the \emph{Natural Cure} topic is low, with an average quality rating of 1.45. In contrast, links returned for queries related to the \emph{Stigma}, \emph{Breastfeeding}, and \emph{Drugs} topics have much higher average quality ratings (4.22, 4.36, and 3.99, respectively). A t-test comparison between \emph{Natural Cure} with each of these topics yields $p < 0.01$. Results for malaria and tuberculosis are similar. Consistent with prior research on H1N1 outbreaks \cite{hill2011natural}, the quality of content that is returned to users varies by topic, especially when we compare \emph{Natural Cure} vs. \emph{Drugs}.


Our analysis could also be used to investigate the quality and volume of information available to individuals with different health information needs.
To get a sense for how much high-quality public health information on natural cures for HIV/AIDS is available, we used the queries most highly associated with the \emph{Natural Cure} and \emph{Drugs} topics on Bing to search authoritative websites on HIV/AIDS. The authoritative websites include the World Health Organization (WHO), the Joint United Nations Programme on HIV/AIDS (UNAIDS), the Center for Disease Control and Prevention (CDC), and the National Institutes of Health (NIH). We posed each of the top 30 queries for \emph{Natural Cure} in turn to each of the aforementioned authoritative websites and noted the number of web pages that were returned on each website for each query. We performed the same actions for the \emph{Drugs} topic. We then reported the average number of web pages available for the 30 queries corresponding to the two topics by website.

We found that on the CDC site, there were an average of $56,705.85$ web pages corresponding to \emph{Natural Cure} (over the 30 queries corresponding to the topic) compared to 258,948.6 for the top 30 queries for \emph{Drugs} $(p = 0.01)$. Similarly, for the NIH site, there were $91,840.8$ and $456,982.3$ $(p = 0.00)$; for the WHO, there were $46,600.1$ and $305,528.2$ $(p = 0.00)$; and for UNAIDS, there were $8,926.5$ and $65,954.0$ $(p = 0.02)$ web pages for \emph{Natural Cure} and \emph{Drugs}, respectively. By this measure, there are consistently fewer high-quality documents for natural cures than for pharmaceutical drugs on authoritative websites.

\section{Discussion and Conclusion}

We have shown that search data, with well-chosen analyses, can provide valuable insights into the health information needs, concerns, and misconceptions of individuals across Africa. Such analyses can complement existing top-down approaches and allow us to narrow the gap in available health data between developing and developed nations.
We conclude with a discussion of the limitations of our techniques as well as implications and next steps for future work.

\vspace{2mm}

\noindent \textbf{Limitations.} There are several limitations to using search data. First, the Bing users we study---and Internet users in general---are not a representative sample of the entire population of Africa. Throughout this work, we have included analyses which give evidence that the data is associated with the diseases at hand. However, since ground-truth data, and especially disaggregated by demographics, is hard to come by due to the health data gap, we are further limited in our ability to measure the representativeness of this data. It is therefore challenging to extrapolate observations obtained through the analysis of search data to the wider population of countries, and the health concerns of entire communities who are not on the web could be overlooked. Still, many findings corroborated previous smaller scale studies using representative samples. 
A second limitation is that the results of this study depend on proprietary data from Bing which can limit the ability for health organizations to extend the research.

Another limitation is the use of imprecise language in search queries, as well as queries in languages other than English. An initial exploration of the Bing query logs showed that many users search for HIV/AIDS, malaria, and tuberculosis by their English names, but it is likely that the filtering method we used still led us to exclude many relevant searches. Furthermore, the excluded searches are more likely to come from regions in which the use of English names is less common, further biasing the data collection. These concerns could be amplified for other illnesses, such as respiratory infections, which have multiple common names in different languages and for which users commonly search for symptoms instead of the disease name itself. A multi-language approach would be necessary to more fully extract all of the information on individuals' health information needs that is captured in search data.

\vspace{2mm}

\noindent \textbf{Practical implications.} Our methods have great potential to inform targeted education efforts in data-sparse regions.
Gender and age impact an individual's chance of contracting HIV/AIDS, malaria, or tuberculosis~\cite{whogender}, and health information needs are often specific to demographic groups and geographic locations.
For these reasons, stakeholders have emphasized the need for gender-responsive and age-responsive programming in resource-constrained regions. Efforts to understand health information needs in developing nations by demographic group have mostly used surveys and interviews, which are limited in their scale \cite{chinahiv,china2,ugandahiv}. Search data allows us to study health needs at a much larger scale.

Search engines themselves could potentially be an effective platform for implementing targeted interventions to improve access to health information. For instance, gender- or age-specific targeted advertisements for health campaigns could be triggered by queries associated with specific health topics. Insights garnered from the analysis of search data could help health organizations prepare material and develop interventions aimed at regions where specific misconceptions are especially common. These interventions could take the form of highlighted high-authority links to discourage misinformation, advertisements for support groups triggered by searches related to stigma, or advertisements for testing clinics for testing-related searches. Recent work has studied the potential to use computational techniques to combat health misinformation on social media \cite{fakecures}.

Search data could also be used to monitor other aspects of public health, for example by providing marketing surveillance for new medications or measuring the impact of public health campaigns. In principle, search engines and health organizations could also work together on case finding, a strategy that directs resources at individuals or groups suspected to be at risk for a particular disease, which is a key strategy in communicable disease outbreak management. Of course, this pursuit would need to be handled with great care, with consideration for the risks and ethics involved.

Finally, as it is detailed and available in real time, search data could be especially valuable for monitoring the impacts of emerging health concerns in developing nations. For instance, noncommunicable diseases such as cancer, cardiovascular disease, and diabetes are of growing concern due to the expansion of the middle class in developing countries and a lack of resources and programs aimed at minimizing their impact \cite{AU}. Since the portion of the population affected by these diseases is likely to have Internet access, search data could play an instrumental role in understanding attitudes about these diseases, implementing interventions to improve access to health information, and highlighting overlooked aspects of the impacts of these diseases.


\bibliographystyle{aaai}
\bibliography{refs}

\appendix

\section{Additional Details on Data and Methodology}

\vspace{2mm}

\noindent \textbf{Data Cleaning.} After generating the initial sets of all queries containing the disease terms (``HIV'' or ``AIDS,'' ``malaria,'' or ``tuberculosis'' or ``TB,'' respectively), removing queries with two or fewer words, and scrubbing the data to remove personal information, including all HIPAA identifiers \cite{hipaa}, we manually examined a sample of 1,000 queries from each data set to check whether they were relevant to the disease. The scrubber we used, Tee anonymizer, extracts and replaces PII with more suitable specific placeholders. For example, an email address gets replaced by the text “emailpii”. There are 13 different types of PII that are replaced including Name, Phone, Address, SSN, CC, and so on. Of these samples, we found that all queries in the malaria data set were related to malaria, but $4.5\%$ of the queries in the HIV/AIDS data set and $16.7\%$ of the queries in the tuberculosis data set were off-topic, containing phrases such as ``tb dresses,'' ``TB Joshua,'' or ``4 TB.'' To improve the quality of the tuberculosis data set, we used these off-topic queries to generate a list of common phrases that we then employed to filter out irrelevant queries from the full tuberculosis data set. After this filtering step, we sampled a fresh set of 1,000 queries and found that only $7.0\%$ were off-topic.

\vspace{2mm}

\noindent \textbf{Languages Used.} One potential source of bias in our data is that the way in which the data were filtered may have caused us to miss relevant queries made in languages other than English. To understand the extent of this potential problem, we examined the set of all search queries made on Bing anywhere in Africa during February 5--11, 2016. We sampled 1,000 random queries from this set and manually determined how many of these queries appeared to be in English. Overall, $22.5\%$ of the queries were in languages other than English. Of the remaining $77.5\%$, all were either in English or could potentially have been in English (i.e., the name of a celebrity or the name of a country). Most non-English queries were in either Arabic, French, or Portuguese, and most non-English searches were concentrated in a few countries. Morocco, Algeria, and Egypt had especially high concentrations of non-English searchers; $49\%$ of queries from Morocco, $55\%$ of queries from Algeria, and $61\%$ of queries from Egypt were not in English.

\vspace{2mm}

\noindent \textbf{User Demographic Distributions.} As reported in the main text, a portion of the queries in our data sets were accompanied by self-reported age and/or gender of the user. Users are identified by an anonymous ID. In this section, we present some analyses to indicate the distribution of these values and associations with population demographic statistics.

We first took all anonymous IDs that are associated with a query in any one of the HIV/AIDS, malaria, or tuberculosis data sets. We look at all users who reported ages between 18 and 50. Recall that this age range corresponds to that considered for our analysis for age and topic usage. (We did not consider individuals who report an age below 18 due to ethical considerations and above 50 due to data sparsity concerns.) We found that this age range account for $90\%$, $86\%$, and $87\%$ of the queries made by users with age available for the HIV/AIDS, malaria, and tuberculosis data sets, respectively.

We further analyzed association of the reported age and gender of the users with population statistics. Specifically, we use the statistics presented by \cite{agestats}, which contains estimated population statistics for different age groups given in 5 year age groups. We took the age ranges between 20 and 49 and consider the fraction of the population in the age ranges 20--24, 25--29, \ldots, 45--49. Likewise, we consider the fraction of users who reported ages in these ranges. We find a correlation coefficient of $\rho = 0.9564 \ [0.6478, 0.9954]$ with $p < 0.01$. Note, however, that this correlation coefficient drops significantly when adding older ages due to the sparsity concerns present in our data.

Furthermore, 54.19\% percent of anonymous IDs which report their gender are men while 45.81\% were women. On the other hand, 50.13 \% of the African population is female while 49.87\% is male \cite{agestats}. This digital gap observed by gender is consistent with findings that women (as well as older age-groups, families in rural regions, low-literacy individuals, and other under-served communities) have less access to Internet \cite{ITU}. Note, however, accurate statistics disaggregated by the above groups is challenging to obtain in many African nations.

\begin{figure}[!ht]
\centering
\includegraphics[width = 4.8cm]{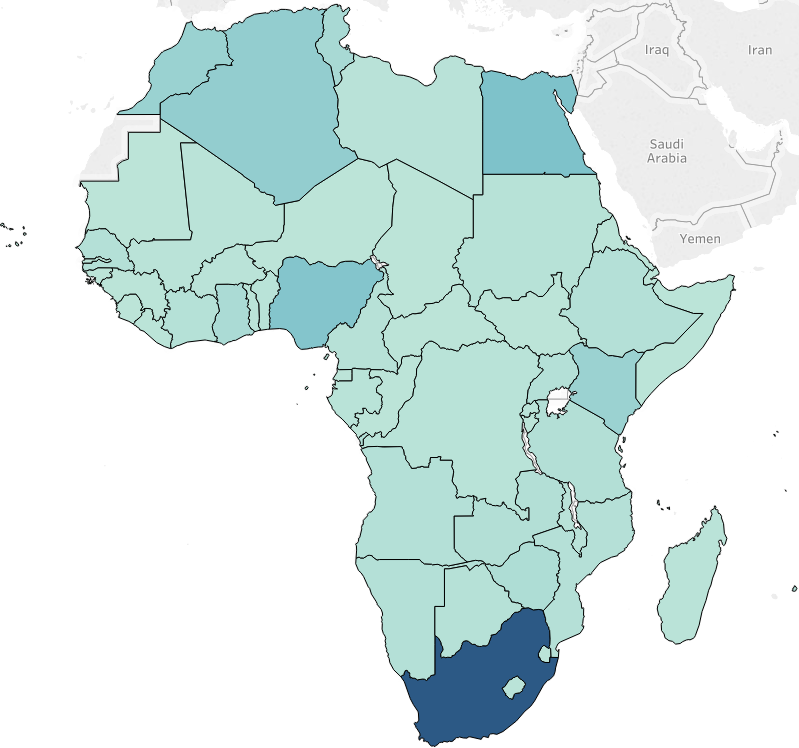} 
\vspace{5mm}
\includegraphics[width = 4.8cm]{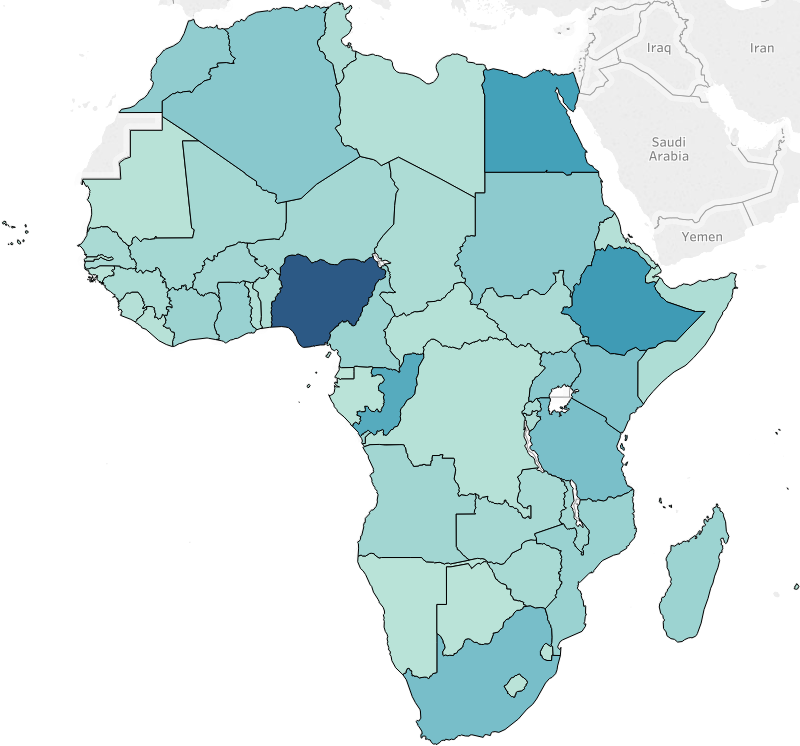}
\caption{Top: Heat map of the total search traffic in each country during the January 2016--June 2018 period. Bottom: Heat map of the population in each country in 2016. \label{fig:penetration}}
\end{figure}

\vspace{2mm}

\noindent \textbf{Coverage of Africa.} Our search data covers all 54 nations of Africa.
Figure \ref{fig:penetration} shows a heat map of the total search traffic during January 2016--June 2017 period by country and a heat map of the 2016 population of each country.
The Spearman correlation coefficient between total search traffic during the January 2016--June 2017 period by country and the 2016 population of each country is $\rho = 0.622 \ [0.591 - 0.651]$ with $p < 0.01$. The correlation between search traffic and Internet penetration is $\rho = 0.574 \ [0.540, 0.606]$ with $p <0.01$.

\begin{table*}[!htbp]
\caption{Additional sample LDA topics for HIV/AIDS, malaria, and tuberculosis with representative words and sample queries.}
\begin{tabular}{p{0.05\textwidth}|p{0.1\textwidth}|p{0.47\textwidth}|p{0.27\textwidth}}
\hline
Disease & Topic & 20 Most Representative Words & Sample Queries from Top 100\\
\hline
\hline
&  Transmission \newline (0.61\%) & sex, oral, penis, infected, man, woman, vagina, person, sucking, risk, contact, pussy, positive, contract, workers, chances, condom, girl, transmitted, sperm
& aids-infected penis \newline hiv cunnilingus \newline sucking penis transmit hiv
\\
 \cline{2-4}
    &   Testing \newline kits \newline (0.69\%)  & test, home, kit, testing, treat, kits, buy, clicks, rapid, tests, tester, app, pharmacy, finger, phone, online, download, free, dischem, price
& hiv home test kit cvs \newline download hiv fingerprint scanner \newline 
hiv kit dischem
\\
\cline{2-4}
 & Gender \newline inequality \newline (0.45\%) & spread, gender, contribute, power, relations, infections, ways, inequality, unequal, infection, discuss, relation, namibia, imbalance, contributes, pdf, poverty, zimbabwe, lead, spreading
& hiv spread gender equality relations \newline unequal hiv infections 
\\
\cline{2-4}
HIV/ \newline AIDS & Healthy \newline lifestyle \newline (0.59\%) & food, positive, people, person, diet, healthy, living, eat, good, patients, patient, medication, nutrition, foods, lifestyle, eating, importance, manage, tea, supplements
 & hiv healthy food diet \newline food insecurity and hiv \newline hiv vitamins and supplements 
\\
\cline{2-4}
& Disease \newline progression \newline (0.48\%) & cd4, count, positive, blood, cells, bebe, winans, low, story, cell, patients, patient, person, white, high, treatment, virus, negative, infection, cd
& hiv cd4 count 500 \newline cd4 cd8 ratio hiv \newline t4 count in hiv
\\
\cline{2-4}
  &     Celebrity \newline gossip \newline (0.95\%) & ***, ***, positive, hiv-positive, status, ***, ***, ***, TRUE, ***, ***, ***, ***, ***, ***, ***, ***, ***, ***, ***
& *** *** hiv-positive \newline *** *** hiv status \newline is *** hiv-positive 
\\
\hline
\hline 
& Prevalence \newline (1.07\%) & map, areas, africa, countries, risk, endemic, kenya, high, area, botswana, cdc, mozambique, affected, namibia, list, african, country, found, zimbabwe, zones
& top malaria countries \newline malaria endemic areas map \newline how europe eliminated malaria
 \\
\cline{2-4}
& Mortality \newline (0.80\%) & mortality, children, rate, nigeria, morbidity, death, impact, due, child, questions, malawi, africa, prevalence, effects, infection, years, maternal, poverty, anaemia, related
  & malaria morbidity and mortality \newline malaria premature \newline child mortality due to malaria
\\
\cline{2-4}
 & Testing kit \newline (1.32\%) & test, rapid, diagnostic, kit, testing, rdt, kits, tests, pf, antigen, sd, positive, results, result, negative, bioline, diagnosis, pan, urine, rdts
 & buy binaxnow malaria combo kit \newline rapid malaria test kit \newline false negative rapid malaria test
\\
\cline{2-4}
Malaria & Parasite \newline (1.54\%) & cycle, life, parasite, 10, icd, plasmodium, diagram, code, stages, history, cdc, mosquito, pdf, parasites, lifecycle, host, describe, explain, transmission
& life cyle malaria \newline cdc malaria life cycle diagram \newline malaria icd
\\
\cline{2-4}
 & Pregnancy \newline (1.11\%) & pregnant, drugs, pregnancy, woman, drug, women, anti, treat, treatment, safe, trimester, weeks, nigeria, list, good, medication, early, medicine, treating
& 36 weeks with malaria \newline malaria in pregnant woman \newline malaria prophylaxes
\\
\cline{2-4}
& Prevention \newline (0.91\%) & prevent, ways, control, prevention, measures, preventing, spread, methods, preventive, method, controlling, prevented, pdf, ddt, reduce, avoid, areas, awareness, community, campaign
& ways of controling malaria \newline malaria preventative measures \newline anti-malaria campaigns 
\\
\hline 
\hline
&  Patient \newline care \newline (1.11\%) & person, people, patients, patient, pictures, food, eat, images, infected, spread, lungs, diet, prevent, picture, hiv, healthy, foods, good, avoid, suffering
& eating healthy with tuberculosis \newline foods to avoid tb \newline food supplements for tb patients
\\
\cline{2-4}
 &   Testing \newline (1.23\%) &  test, positive, skin, pictures, negative, results, blood, gold, quantiferon, sputum, tests, testing, lam, result, urine, diagnosis, diagnostic, pcr, FALSE, rapid
& quantiferon gold tb \newline capilla tb test \newline cdc tb skin test reading
\\
\cline{2-4}
TB &  Prevention (0.95\%) & stop, strategy, end, dots, partnership, life, reach, meaning, quality, program, cycle, order, logo, treatment, price, application, key, wave, strategies, nigeria
& stop tb partnership \newline tuberculosis life cycle \newline tb reach wave 6
\\
\cline{2-4}
&   HIV \newline co-infection (0.94\%) & hiv, coinfection, research, nigeria, co-infection, job, description, 2017, patients, positive, jobs, aids, related, tvoroyri, solution, tanzania, project, children, eradication, field
&  tb-hiv co-infection \newline tuberculosis description \newline tb coordinator job description
\\
\cline{2-4}
&  National \newline programs   (1.23\%) & national, control, hiv, plan, program, leprosy, health, strategic, programme, infection, policy, guidelines, prevention, kenya, sti, ministry, aids, manual, training, nigeria
& tb crisis plan \newline tanzania strategic plan tb \newline tb helpline ghana
 \\
\cline{2-4}
 & Global \newline programs \newline (0.92\%) & hiv, aids, malaria, global, fund, fight, health, project, program, services, nigeria, funding, diseases, impact, integration, challenges, lesotho, grant, call, cholera
& global fund fight aids tuberculosis malaria \newline tb malaria cholera meningitis  
\\
\hline
\end{tabular}\label{table:topics}
\end{table*}

Building on the results from Figure 1, we run a multiple linear regression with the HIV prevalence rate as the dependent variable and the fraction of searches associated with the disease, Internet penetration, percent of population in urban settings, population, and GDP as the explanatory variables. We find that the fraction of searches containing the disease name explain some of the variance in disease prevalence even after controlling for the other explanatory variables.

\section{Additional Details on the Topics}

To give a better sense of the breadth of topics output by LDA, we provide an additional six example topics for each disease in Table \ref{table:topics}. These topics were again manually selected by the authors to show the wide range of themes that emerge. As in the main text, the labels in the second column are provided by the authors, the third column displays representative words for each topic, and the fourth column contains a randomly selected sample of the 100 most representative queries. Before running LDA, we scrubbed the data to ensure that HIPAA identifiers were not included. We then manually scrubbed the output of LDA to further remove identifiers, such as celebrity names, which have been redacted here.

\section{Analyses of the Popularity of Topics}

\vspace{2mm}

\begin{figure}[!ht]
\centering
\includegraphics[scale = 0.09]{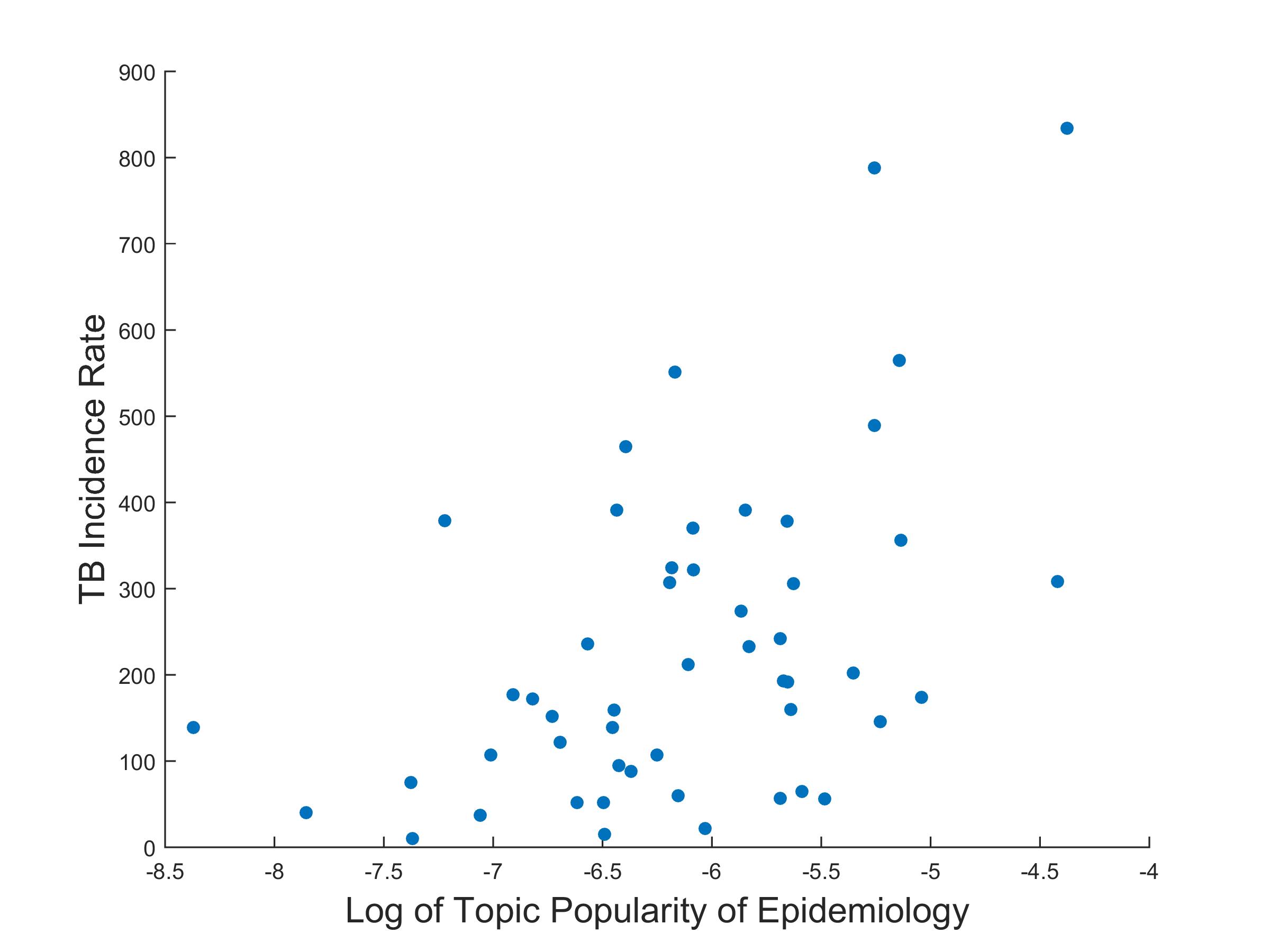}
\caption{Popularity of the \emph{Epidemiology} topic vs. tuberculosis incidence. \label{fig:tbtopics}}
\end{figure}

\noindent \textbf{Topic Popularity by Country.} In the main paper, we discuss the association between the popularity of the \emph{Stigma} topic for HIV/AIDS in a country and that country's disease prevalence. We ran similar tests on the six topics included in the table in the main body for the malaria and tuberculosis data sets. We find a significant (multi-test corrected) relationship between the popularity of the \emph{Epidemiology} topic for tuberculosis and the tuberculosis incidence rate ($r = 0.509$ multi-test corrected $p < 0.01$). See Figure~\ref{fig:tbtopics}.\footnote{As in the main text, relationships are considered significant if they pass the Benjamini-Hochberg false discovery rate (multi-test correction) with $\alpha = 0.5$.}

\vspace{2mm}

\noindent \textbf{Topic Popularity by User Demographics.} As we did for HIV/AIDS, we looked at the topic popularity of the six topics of interest from the topic table in the main paper. Tables \ref{table:malariafull} and \ref{table:tbfull} list the correlation coefficients, where $**$ indicates a p-value of less than 0.01 and $*$ indicates a p-value of less than 0.05.
As we saw in the HIV/AIDS data set, women and individuals in the 25--34 age group expressed relatively more interest in topics related to pregnancy, breastfeeding, and family care compared to men in the malaria data set. Additionally, for both the malaria and tuberculosis data sets, users in the 25--34 age group were relatively more interested in symptom-related topics. This is consistent with the literature that nearly half of all new HIV infections occur among the 15--24 age group. The 25--34 age group corresponds to the time-frame where these infections may progress significantly, and even develop to AIDS.

\begin{table}[!ht]
\caption{Relative topic popularity by user demo. for malaria. \label{table:malariafull}}
\begin{tabular}{p{0.08\textwidth}|p{0.071\textwidth}|p{0.071\textwidth}|p{0.071\textwidth}|p{0.071\textwidth}}
\hline
& Women & Ages 18--24 & Ages 25--34 & Ages 35--49 \\
\hline
\hline
Drug & 0.006 & -0.068** & -0.008 & 0.031 \\
Natl Cure & 0.027 & -0.051** & -0.014 & 0.051** \\
Breast. & 0.073** & -0.089** & 0.060** & 0.031 \\
Epidem. & -0.084** & -0.002 & -0.042* & -0.005 \\
Diagnosis & -0.065** & 0.032 & 0.071** & -0.058** \\
Symptoms & 0.055** & 0.001 & 0.062** & -0.019
\\
\hline
\end{tabular}\label{table:mtopicpopularity}
\end{table}

\begin{table}[!ht]
\caption{Relative topic popularity by user demo. for TB.\label{table:tbfull}}
\begin{tabular}{p{0.08\textwidth}|p{0.071\textwidth}|p{0.071\textwidth}|p{0.071\textwidth}|p{0.071\textwidth}}
\hline
& Women & Ages 18--24 & Ages 25--34 & Ages 35--49 \\
\hline
\hline
Epidm. & 0.006 & 0.037* & -0.078** & 0.007 \\
Drug Res. & 0.003 & 0.007 & -0.027 & 0.002 \\
Diagnosis & -0.043* & 0.004* & -0.011* & 0.001 \\
Symptoms & 0.090** & 0.014 & 0.074** & -0.042 \\
Side-effe. & 0.027 & 0.028 & -0.004 & -0.016 \\
Natl Cure & 0.018 & -0.020 & -0.016 & 0.022
\\
\hline
\end{tabular}\label{table:tbtopicpopularity}
\end{table}

We looked at the topic popularity of the six topics of interest from the topic table in the main paper. Women and individuals in the 25--34 age group expressed relatively more interest in topics related to pregnancy, breastfeeding, and family care compared to men in the malaria data set. Additionally, for both the malaria and tuberculosis data sets, users in the 25--34 age group were relatively more interested in symptom-related topics.

\section{Additional Details on User Behavior and Quality of Results}

\subsection{User Behavior}

\begin{figure*}[!ht]
\centering
\vspace{4mm}
\includegraphics[width = 5.85cm ]{clickcount.jpg}
\includegraphics[width = 5.85cm]{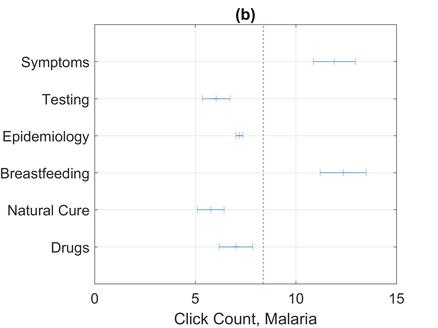}
\includegraphics[width = 5.85cm]{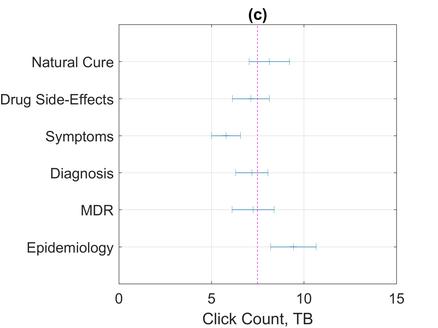}\\
\vspace{4mm}
\includegraphics[width = 5.85cm]{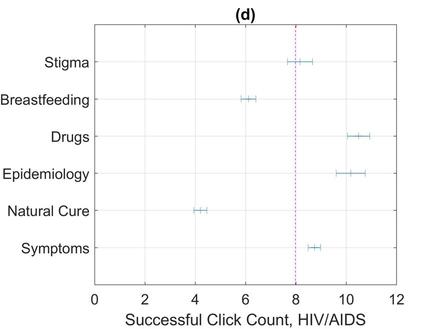}
\includegraphics[width = 5.85cm]{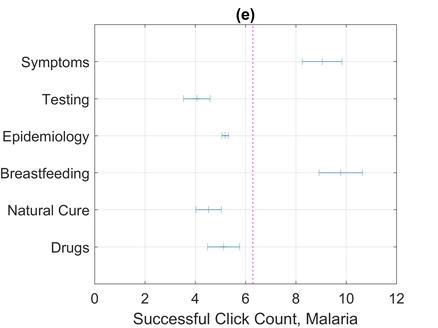}
\includegraphics[width = 5.85cm]{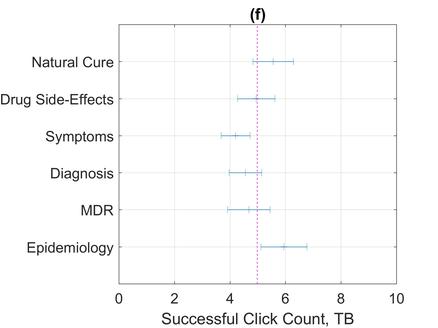}\\
\vspace{4mm}
\includegraphics[width = 5.85cm]{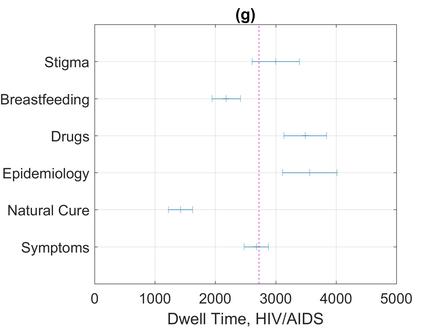}
\includegraphics[width = 5.85cm]{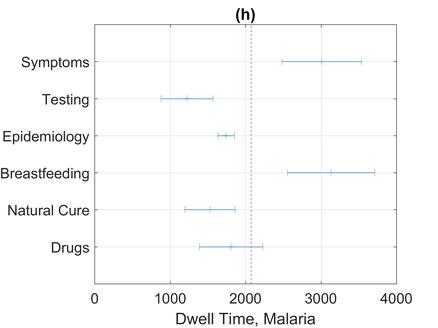}
\includegraphics[width = 5.85cm]{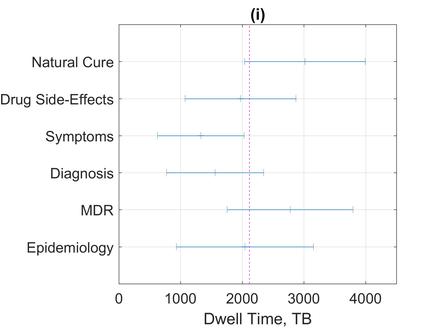}\\
\vspace{4mm}
\includegraphics[width = 5.85cm]{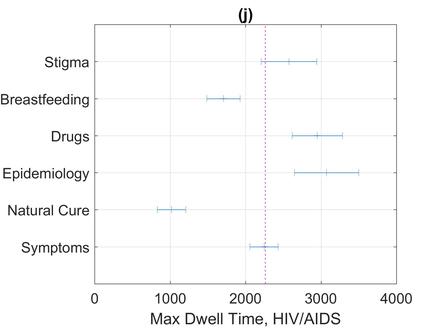}
\includegraphics[width = 5.85cm]{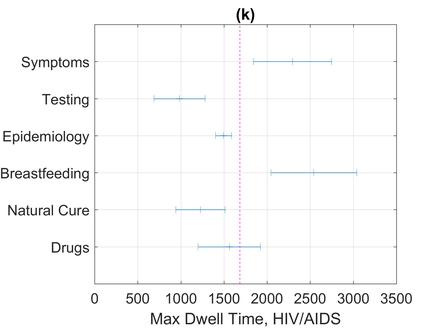}
\includegraphics[width = 5.85cm]{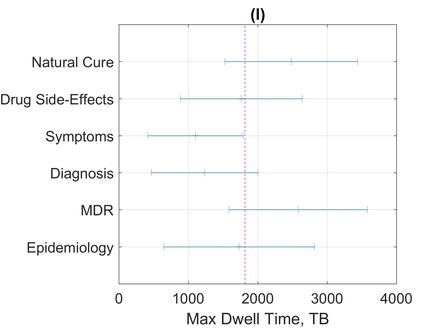}
\vspace{4mm}
\caption{Rows, from top to bottom: average click count, successful click count, dwell time, and maximum dwell time for queries associated with different topics. Columns, from left to right: HIV/AIDS, malaria, and tuberculosis data sets. The vertical lines represent the mean values across topic. \label{fig:users}}
\end{figure*}

We examined whether user behavior varies across different topics for the HIV/AIDS, malaria, and tuberculosis data sets. We used four popular metrics in the information retrieval literature: dwell time, maximum dwell time, click count, and successful click count. Dwell time and click count are discussed in the main text.
\emph{Maximum dwell time} measures the maximum amount of time that a user spends on any link that is followed. 
\emph{Successful click count} is the total number of links the user clicks that have a dwell time of at least 30 seconds. We use the same methodology described in the main text and the regression coefficients to report the average values.  
Results are shown in Figure \ref{fig:users}.  Note that for the HIV/AIDS and malaria data sets, users issuing queries associated with \emph{Natural cure} exhibited relatively low activity (by all four metrics) compared to many of the other topics of interest; this is not true for the tuberculosis data set.

\subsection{Quality of Content}

To measure the quality of the links presented by topic, we examined the set of links returned in the first position for the thirty most representative queries for each of the six topics from the malaria and tuberculosis data sets that appear in the table in the main text, in the same way described in the main text for the HIV/AIDS data set. 
Each link was evaluated by three research assistants, each of whom has graduate-level training in medicine or public health, and at least one of whom specializes in the corresponding disease.
In all three cases, the research assistants were asked to assess the relevance, accuracy, and objectiveness of the links returned. 
  In particular, the research assistants were presented with the following questions.
\begin{itemize}
\item \emph{Relevance:} How comprehensive and complete is the information provided on the website and does it appear to provide the right level of detail? Is the URL related to the disease? 
\item \emph{Accuracy:} Are sources of information properly identified, and are there any glaring omissions or misinformation?
\item \emph{Objectiveness:} Does the information presented appear in an objective manner without political, cultural, religious, or institutional bias?
\end{itemize} 
The research assistants were asked to assign a single rating for each link on a scale from $1$ to $5$, with 1 equal to bad quality and 5 equal to high quality. Values were defined as:
\begin{enumerate}
\item Bad quality: several serious issues concerning all three of relevance, accuracy, and objectiveness
\item Subpar quality: several serious issues covering at least two of relevance, accuracy, or objectiveness
\item Mediocre quality: several issues concerning relevance, accuracy, or objectiveness
\item Good quality: mostly relevant, accurate, and objective, with a few small issues

\item High quality: very relevant, accurate, and objective
\end{enumerate}
Note that the research assistants were not asked to take into account the website design, interface, usability, or other metrics unrelated to content.

\begin{figure*}[!ht]
\centering
\includegraphics[width = 6cm]{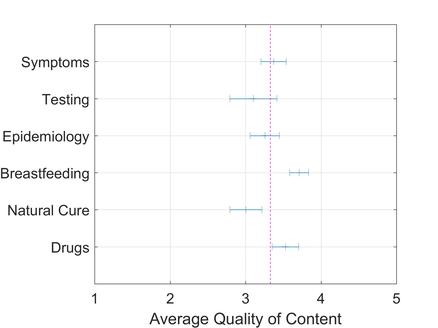}
\includegraphics[width = 6cm]{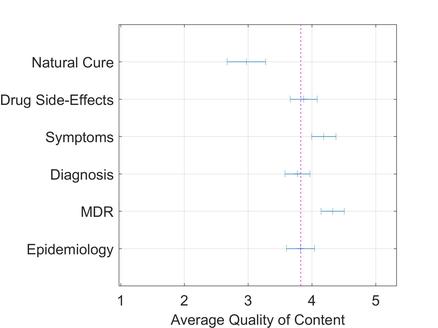}
\caption{Average quality of content of webpages returned to users for the 30 queries most strongly associated with the six malaria (left) and tuberculosis (right) topics.  \label{fig:qualityofcontent}}
\end{figure*}

Consistent with the observations for the HIV/AIDS data set, for the tuberculosis data set, the quality of content returned to users was, on average, rated lower for queries related to the \emph{Natural cure} topic than for other topics of interest. In contrast, for the malaria data set, the quality of links returned was indistinguishable among queries related to the \emph{Natural cure}, \emph{Epidemiology}, \emph{Testing}, and \emph{Symptoms} topics.

\end{document}